\documentclass[reprint, amsmath, amssymb, aps, twocolumn, superscriptaddress, floatfix, pra, hidelinks]{revtex4-1}

\usepackage[utf8]{inputenc}
\usepackage{graphicx}
\usepackage{hyperref}
\usepackage{dcolumn}
\usepackage{bm}
\usepackage{tcolorbox}
\usepackage{lipsum} % For dummy text, can be removed
\newcommand{\PRLSec}[1]{\textit{#1}---}
\usepackage{float}

\usepackage{cleveref}

\begin{document}

%###########################################################################

\title{Optical Tautochrone and Squeezing Dynamics in Nonuniform Lattices}

\author{Ioannis Kiorpelidis}
\email{gkiorpelidis@iesl.forth.gr}
\affiliation{ITCP-Department of Physics, University of Crete, 71003  Heraklion, Greece}
\affiliation{Institute of Electronic Structure and Laser (IESL), FORTH, 71110 Heraklion, Greece}

\author{Matthias Heinrich}
\affiliation{Institute for Physics, University of Rostock, Rostock, Germany}

\author{Alexander Szameit}
\affiliation{Institute for Physics, University of Rostock, Rostock, Germany}

\author{Georgios A. Siviloglou}
\email{g.siviloglou@physics.uoc.gr}
\affiliation{ITCP-Department of Physics, University of Crete, 71003  Heraklion, Greece}
\affiliation{Institute of Electronic Structure and Laser (IESL), FORTH, 71110 Heraklion, Greece}
\affiliation{Department of Physics, Southern University of Science and Technology, Shenzhen 518055, China} 
\affiliation{International Shenzhen Quantum Academy, Shenzhen 518055, China}

\author{Konstantinos G. Makris}
\email{makris@physics.uoc.gr}
\affiliation{ITCP-Department of Physics, University of Crete, 71003  Heraklion, Greece}
\affiliation{Institute of Electronic Structure and Laser (IESL), FORTH, 71110 Heraklion, Greece}

%###########################################################################

\begin{abstract}
We present exact analogies between the tautochrone problem of classical mechanics and the squeezed states of quantum optics to optical lattices.
Both phenomena emerge in the same physical system, that of waveguide arrays with nonuniform couplings. 
Extension to two dimensions yields Lissajous-type trajectories and multidirectional tautochrone focusing. 
Furthermore, we investigate the impact of Kerr nonlinearity and show that it determines the diffraction behavior, namely coherent-state-like or squeezed propagation.
These quantum inspired classical lattices highlight the role of the coupling coefficients  to beam engineering and light control in complex media.
\end{abstract}

\maketitle

%###########################################################################

\label{Section1}
\PRLSec{Introduction}One of the most celebrated problems in the history of physics is the tautochrone \cite{Simmons1972}. Its origins trace back to the 17th century, when Christiaan Huygens demonstrated that a particle sliding  under gravity along a cycloid curve and without friction reaches the lowest point in the same time irrespective of its initial position, provided that it starts from rest \cite{Huygens}. Notably, the cycloid is also the solution to the brachistochrone problem, namely the challenge posed by Johann Bernoulli of finding the curve of fastest descent between two points under gravity \cite{Boyer1991}. 
The brachistochrone problem has found applications in a wide range of areas, from the study of time-optimal processes in quantum dynamics and control theory \cite{Carlini2006}, to modern shortcuts to adiabaticity methods \cite{Guery2019}.

In a seemingly unrelated direction, a nonlinear optics platform that has been extensively investigated in recent years, as it offers a fertile ground for demonstrating  diverse physical phenomena, is that of photonic lattices \cite{Christodoulides1988, Christodoulides2003}, i.e., arrays of evanescently coupled waveguides. The dynamics in linear photonic lattices is controlled by two parameters: the on-site potential of each waveguide and the coupling strength between neighboring channels \cite{Silberberg2008}. When the on-site potential is modulated across the lattice, and light propagates without nonlinear interactions, these optical systems have enabled the observation of several hallmark effects, such as Bloch oscillations \cite{Peschel1998, Hartmann2004}, Anderson localization \cite{Thompson2010,Martin2011}, Rabi oscillations \cite{Shandarova2009}, and accelerating Wannier–Stark states \cite{Ganainy2011}, among others. Such photonic lattices have also been employed to explore topological phases of light and edge-state transport \cite{Rechtsman2013, Biesenthal2022}. When nonlinear effects are introduced, additional wave phenomena have been demonstrated, for instance soliton dynamics \cite{Eisenberg1998}, filament formation \cite{Belec2012}, and Bloch oscillations due to nonlinearity \cite{Driben2017}.

Additionally, the modulation of the coupling coefficients across the lattice provides an extra degree of freedom. A prominent example of that type is the Glauber-Fock lattice \cite{Leija2010, Keil2011, Lara2011, Longhi2013}, where coherent and displaced Fock states emerge.
Another well-studied configuration is the so-called \( J_x \) lattice \cite{Gordon2004, Cessa2023, Nikolopoulos2004, Rai2021, Wolterink2023}, in which the coupling strength between waveguides follows a parabolic law, leading to oscillatory beam dynamics \cite{Longhi2010}, that in turn has enabled the transfer of light between two distant  sites \cite{ Chremmos2012, Bellec2012, Szameit2013, Chapman2016}. It is noted that this parabolic coupling profile was originally proposed in engineered spin chains, where it resulted in the perfect state transfer of an initial excitation at one end of the spin network to the opposite end, at a prescribed time \cite{Landahl2004}.

\begin{figure}
\begin{center}
\includegraphics[width=1\columnwidth]{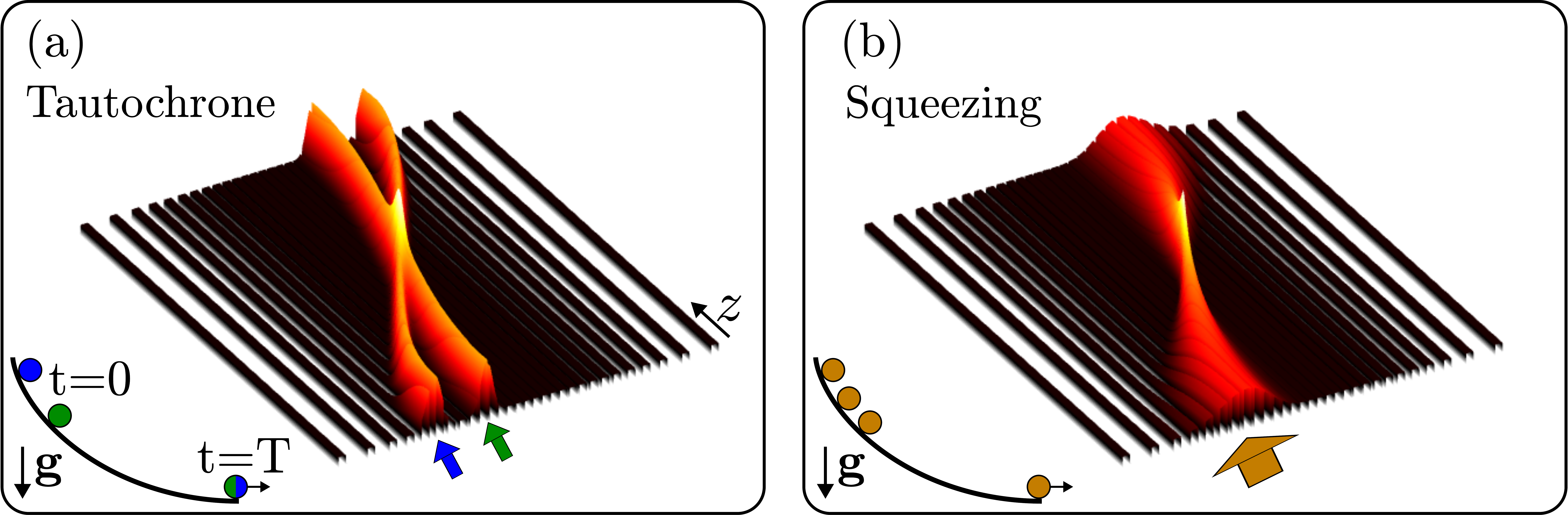}
\caption
{
{Optical tautochrone  and squeezing dynamics in non-uniform photonic lattices.}
(a) Schematic illustration of the tautochrone effect: two Gaussian inputs are launched at different positions in the lattice and focus after a fixed propagation distance $z$. Inset: classical tautochrone, where particles sliding along a cycloid under gravity reach the bottom simultaneously, regardless of their starting positions.
(b) Schematic illustration of squeezing dynamics: a broad Gaussian input undergoes periodic squeezing during propagation. Inset: the wave packet is viewed as a collection of particles that all arrive together at the bottom of the curve, resulting in compression.
}
\label{fig1}
\end{center}
\end{figure}

In this Letter, we provide a class of photonic lattices with nonuniform couplings where the oscillatory trajectories of the beams  resemble two different physical phenomena. First, these oscillations yield a discrete analog of the tautochrone phenomenon of classical mechanics: beams launched from different sites focus at the same location after a fixed propagation distance [see Fig.~\ref{fig1}(a)]. 
Second, the diffraction of an individual beam follows the dynamics of squeezed states of quantum optics \cite{Sipe2013, Kolobov1999, Breitenbach1997, Orzel2001, Aasi2013}: it can either remain shape preserving  or exhibit squeezed evolution with periodic width oscillations [see Fig.~\ref{fig1}(b)]. Moreover, using the Wigner function formalism, we identify the regions of parameter space that support coherent-state-like propagation. Finally,  we demonstrate that Kerr nonlinearity enables switching between the coherent-state-like and squeezed regimes.

%#################################################################################################

\label{Section2}
\PRLSec{The optical tautochrone effect}We consider the propagation of light in single-mode waveguides that form a nonuniform optical lattice with only nearest-neighbor couplings.
The evolution of the optical field is governed by the paraxial coupled mode equations:
$
    i \,\frac{d\boldsymbol{\psi}}{dz} = \mathbf{H} \boldsymbol{\psi},
$
where \( \boldsymbol{\psi} = [\psi_1, \psi_2, \ldots, \psi_N]^T \) and \( \psi_i \) is the amplitude of the electric field envelope in the \( i^{\mathrm{th}} \) waveguide. Coupling between neighboring sites is encoded in the off-diagonal terms of the Hamiltonian \( \mathbf{H} \). In the one-dimensional case (1D), the evolution equations read,
\begin{align}
    i \tfrac{d\psi_1}{dz} &= -J_1 \psi_2, \\
    i \tfrac{d\psi_n}{dz} &= -J_{n-1} \psi_{n-1} - J_n \psi_{n+1}, \quad n = 2,\ldots,N-1, \\
    i \tfrac{d\psi_N}{dz} &= -J_{N-1} \psi_{N-1},
\end{align}
with \( J_n \) being the hopping amplitude between waveguides \( n \) and \( n+1 \).
To construct the lattice that supports the  tautochrone effect---namely, the convergence of beams launched from different positions---we require that each beam undergoes oscillatory motion during propagation. 
By employing the Ehrenfest theorem, in the continuous limit, we find that the coupling profile [details are provided in Supplemental Material (SM) \cite{Supplemental}]
\begin{equation}
    J_n = \omega \sqrt{C^2 - \left(n - \tfrac{N}{2}\right)^2}, \quad n = 1,2,\ldots,N-1,
    \label{couplings}
\end{equation}
gives rise to such an oscillatory evolution around the lattice midpoint, where \( \omega \) is the spatial oscillation frequency and \( C \geq N/2 \) controls the degree of inhomogeneity.  
The validity of the continuous approximation in our discrete system is examined in  SM \cite{Supplemental}.
It is  noted here that the coupling profile given in Eq.~\eqref{couplings} reduces to that of the $J_x$ lattice in the limit $C=N/2$, a system which has an equidistant eigenvalue spectrum (presented in the SM \cite{Supplemental}).
Such equidistant spectra are a hallmark of engineered spin chains \cite{Landahl2004} and dual cavity arrays \cite{Meher2017}, where they enable the perfect state transfer of excitations.

\begin{figure*}
\begin{center}
\includegraphics[width=2\columnwidth]{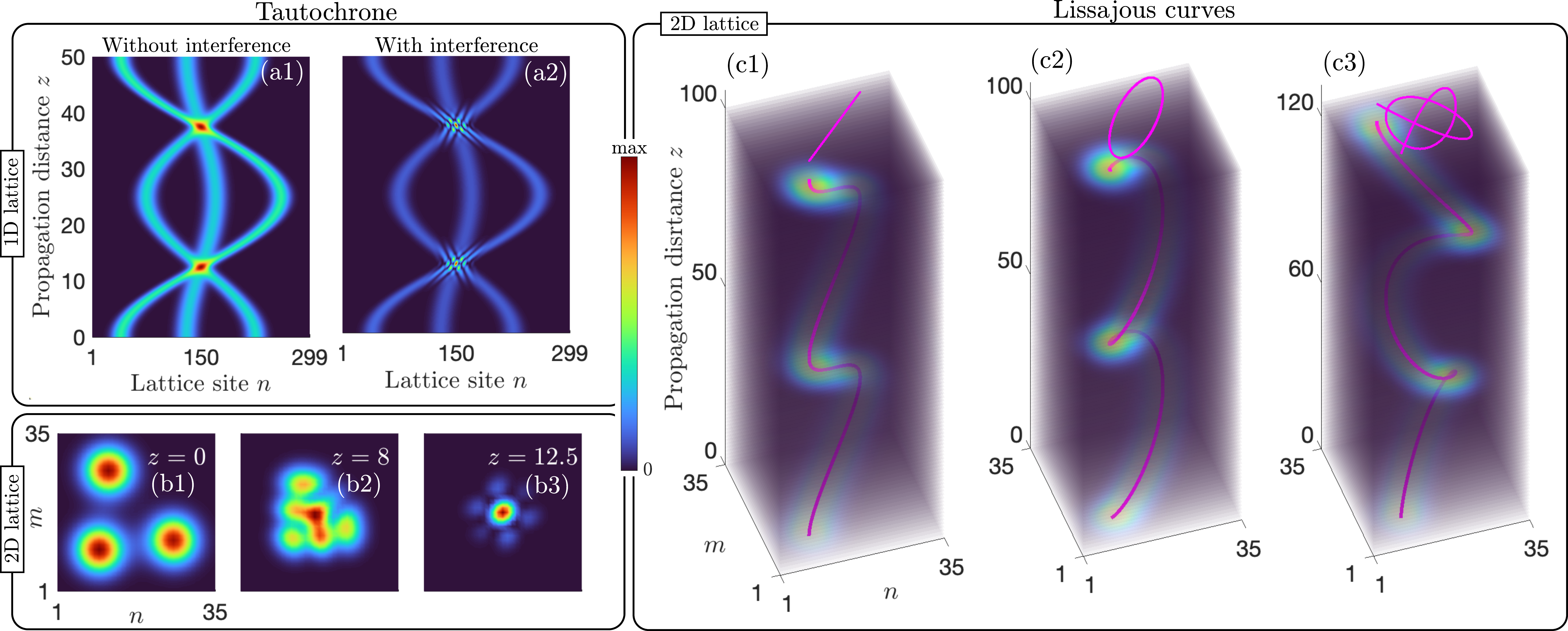}
\caption
{
{Tautochrone effect and oscillatory dynamics in optical lattices.}
(a1) Intensity evolution of {three noninterfering Gaussian wave packets} with zero initial momentum ($p_0=0$) in a one-dimensional lattice with $N=299$ sites. 
We choose the following parameters in Eq.~\eqref{couplings}: \( C = \tfrac{4N}{7} \) and \( \omega = \tfrac{2\pi}{100} \).
The first wave packet is centered at site \( n_0 = 40 \) with width \( w_0 = 11.4 \), the second at \( n_0 = 130 \) with \( w_0 = 13 \), and the third at \( n_0 = 225 \) with \( w_0 = 12.4 \).
(a2) Same 1D setup as in (a1) but with the {three beams interfering}; we display the site-resolved intensity $I_n(z) = |\psi_n(z)|^2$.
(b1)–(b3) Extension to a {two-dimensional lattice} ($35 \times 35$ sites). Shown is the intensity evolution of {three interfering wave packets}. The first wave packet is centered at \( (n_0 = 10, m_0 = 10) \) with equal widths \( w_0 = 4.3 \) in both directions. The second wave packet is centered at \( (n_0 = 26, m_0 = 12) \) 
with widths \( w_{0,x} = 4.3 \) along the \( x \) direction and \( w_{0,y} = 4.4 \) along the \( y \) direction.
The third wave packet is centered at \( (n_0 = 12, m_0 = 27) \), with widths \( w_{0,x} = 4.4 \) and \( w_{0,y} = 4.2 \).
All wave packets have zero initial momentum.
(c1)–(c3) Evolution of a {single wave packet} in a 2D lattice, illustrating the effect of momentum and frequency variations: (c1) zero initial momentum, equal frequencies ($\omega_x = \omega_y=2\pi/100$);
(c2) nonzero momentum in $y$-direction ($p_y = -0.4$), equal frequencies $\omega_x = \omega_y=2\pi/100$; (c3) zero momentum, unequal frequencies $\omega_x=2\pi/100$ and $\omega_y=4\omega_x/5$.
}
\label{fig2}
\end{center}
\end{figure*}

Figures~\ref{fig2}(a1) and \ref{fig2}(a2) illustrate the oscillatory dynamics and the tautochrone phenomenon in a one-dimensional lattice composed of \( N = 299 \) sites. 
The initial state consists of three beams, each centered at a different site. Each beam is Gaussian-shaped and given by
\begin{equation}
    \psi_n = \frac{1}{\sqrt{2\pi w_0^2}} \exp\left[-\frac{(n - n_0)^2}{2w_0^2}\right] \exp( i p_0 n),
    \label{Gaussian}
\end{equation}
where $n_0$ is the center, $w_0$ the width, and $p_0$ the momentum of the wave packet, respectively.
In this example, we set \( p_0 = 0 \); the influence of nonzero momentum will be discussed below. Figure~\ref{fig2}(a1) shows the evolution in a configuration where the three beams do not interfere.  
In this case, we display the total intensity which is the incoherent sum of the three individual intensity patterns.
Noninterfering propagation can be experimentally achieved by using lasers operating at slightly different wavelengths.
We first note that each beam exhibits oscillatory motion around the lattice center ($n = 150$). Specifically, the center-of-mass trajectory $x(z)$ for a beam with zero initial momentum is governed by (see  SM \cite{Supplemental}):
$
x(z) = \tilde{x} + (n_0 - \tilde{x})\cos(2\omega z)
$
where $\tilde{x}$ is the lattice midpoint and $n_0$ is the initial launch site. Crucially, this evolution demonstrates that the oscillation frequency remains independent of the inhomogeneity parameter $C$. Moreover, after a propagation distance $z = \pi/(4\omega)$, the displacement from the center vanishes for all trajectories regardless of their starting position $n_0$: all components focus simultaneously at the lattice midpoint, demonstrating the optical analog of the tautochrone effect.
Figure~2(a2) presents the corresponding evolution in a configuration where the three beams interfere during propagation. Despite this interference, the collective motion still exhibits the hallmark of the tautochrone effect.

We now turn to the extension of the latter results to the corresponding two-dimensional (2D) lattice.
We construct this 2D configuration by implementing the 1D coupling distribution along both orthogonal directions, thus forming a square array with inhomogeneous couplings along both axes.
Figures~\ref{fig2}(b1)-(b3) illustrate the evolution of three interfering beams in this 2D setting. As shown in panel (b3), the beams focus simultaneously at the center of the lattice, demonstrating the two-dimensional manifestation of the optical tautochrone phenomenon.
Furthermore, the trajectory of each beam center traces out a Lissajous curve. This is illustrated in Figs.~\ref{fig2}(c1)–(c3). In particular, Fig.~\ref{fig2}(c1) shows the evolution of a beam with zero initial momentum in both directions; in this case, the  mean position of the beam follows a straight-line path. When a nonzero momentum is introduced along one direction, the resulting trajectory becomes elliptical, as shown in Fig.~\ref{fig2}(c2). Finally, in Fig.~\ref{fig2}(c3), the initial velocities are again zero in both directions, but the oscillation frequencies differ along the two axes, producing a curved trajectory. 
Let us note that, in all the results presented so far, the width \( w_0 \) of each beam was chosen such that the beam maintains its shape throughout the propagation. In the following, we investigate how varying \( w_0 \) influences the dynamics.

%###########################################################################

\begin{figure*}
\begin{center}
\includegraphics[width=2\columnwidth]{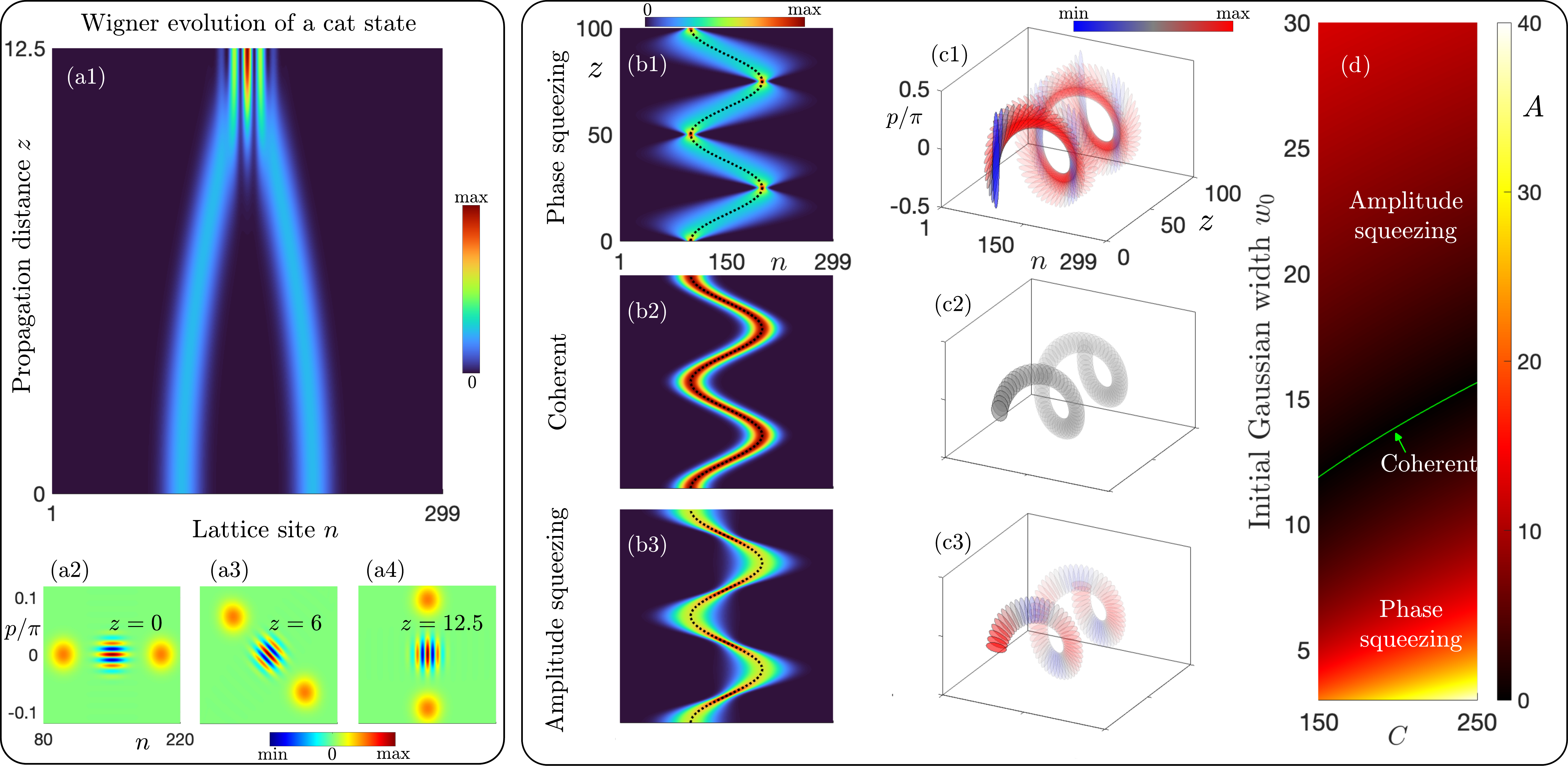}
\caption
{
{Coherent-state-like and squeezing evolution}.
(a1) Evolution of two Gaussian wave packets forming a cat-like state. The first (second) wave packet is centered at \(n_0 = 100\) (\(n_0 = 200\)) with equal widths \(w = 12.8\). Lattice parameters: \( C = \tfrac{4N}{7} \) and \( \omega = \tfrac{2\pi}{100} \).
(a2)–(a4) Discrete Wigner distribution of the state in (a1) at propagation distances \(z = 0\), \(z = 8\), and \(z = 12.5\). 
(b1)–(b3) Evolution of the wave packet magnitude \(|\psi_n(z)|\) for a Gaussian input centered at \(n_0 = 100\), with \(p = 0\) and initial widths \(w_0 = 4.8\), \(12.8\), and \(20.8\), respectively. 
(c1)–(c3) Evolution of the $W(n,p)=10^{-3}$ contour of the Wigner function for the same parameters as in (b1)–(b3). 
(d) Amplitude of width oscillations \(A\) as a function of \(w_0\) and \(C\).
}
\label{fig3}
\end{center}
\end{figure*}

\PRLSec{Coherent-state-like and squeezing evolution}
\label{Section3}The dynamics observed in the under study lattice, mirror those of a quantum harmonic oscillator. 
This analogy is demonstrated in Figs.~\ref{fig3}(a1)–(a4), which show the evolution of a state that is composed of two Gaussian beams symmetrically positioned around the lattice center; this state  constitutes a classical analogue of a quantum cat state.
In particular, Fig.~\ref{fig3}(a1) illustrates the evolution of the site-resolved intensity $|\psi_n(z)|^2$, while Figs.~\ref{fig3}(a2)–(a4) present the corresponding evolution of the discrete Wigner distribution \cite{Blackledget2006}.
We note here that for a one-dimensional lattice, the discrete Wigner function is defined as~
\begin{equation}
    W(n,p) = \sum_{m} \psi^*_{n+m}\,\psi_{n-m}\, e^{2 i p m},
    \label{eqWigner}
\end{equation}
where \(n\) denotes the lattice site, \(p\) is the conjugate momentum, and the summation runs over the integer displacements $m$ that ensure the indices $n \pm m$ remain within the physical boundaries of the $N$-site lattice.
As is shown in Figs.~\ref{fig3}(a2)–(a4), with increasing propagation distance \(z\), the Wigner distribution undergoes a rotation in the \((n,p)\) plane, analogous to the phase-space evolution of a quantum harmonic oscillator.

This analogy between the classical lattice dynamics and the quantum harmonic oscillator motivates us to explore whether states resembling coherent and squeezed states can be supported within the lattice. 
To this end, we examine how the initial beam width \(w_0\) influences the dynamics.
In Figs.~\ref{fig3}(b1)–(b3) we present the evolution of the  magnitude $|\psi_n(z)|$ for three different choices of \( w_0 \):
in Fig.~\ref{fig3}(b1), the initial condition corresponds to a narrow beam, i.e., a small \( w_0 \).
Two key features are evident: (i) the packet continues to undergo oscillatory motion (the black dashed line shows the position of the Gaussian center), and (ii) its shape periodically broadens and compresses as it propagates through the lattice. That is, its width varies with the propagation distance \( z \), i.e., \( w = w(z) \), with initial value \( w(0) = w_0 \).
This situation is analogous to \emph{phase squeezing} in quantum optics, where the phase uncertainty is reduced at the expense of enhanced fluctuations in the amplitude quadrature.
A similar broadening and compression of the shape is observed when \( w_0 \) is large, as shown in Fig.~\ref{fig3}(b3).
In the latter regime, the evolution corresponds to \emph{amplitude squeezing}, in which the amplitude noise is suppressed while the phase fluctuations increase. 
Moreover, there exists a particular initial width \( w_0 \), used in Fig.~\ref{fig3}(b2), for which \( w(z) = w_0 \) along the whole trajectory. In the latter situation the beam preserves its width during propagation; the propagation is coherent-state-like.
Furthermore, Figs.~\ref{fig3}(c1)–(c3) show the evolution of the Wigner distribution  [Eq.~\eqref{eqWigner}] corresponding to the beams depicted in Figs.~\ref{fig3}(b1–(b3).
In particular, each panel depicts how a contour of the Wigner distribution evolves with the propagation distance~$z$. 
In Fig.~\ref{fig3}(c1), the contour is an ellipse elongated along the $p$ axis, while in Fig.~\ref{fig3}(c3) it is an ellipse elongated along the $x$ axis.
By contrast, in Fig.~\ref{fig3}(c2) the contour is a circle.
The background shading indicates the projection of
the contour onto the x-axis, corresponding to the evolution of $w(z)$.
In Figs.~\ref{fig3}(b1) and \ref{fig3}(b3) the background shading varies with~$z$, indicating periodic squeezing and broadening. 
In Fig.~\ref{fig3}(b2), even though the circular contour also rotates with~$z$, its projection remains unchanged, and the background shading does not vary, reflecting coherent-state-like evolution.  It is stressed here that in  SM \cite{Supplemental} we further explore the tautochrone and squeezing effects by investigating the role of the coupling inhomogeneity $C$, by demonstrating the focusing of random initial states, and by examining the impact of coupling disorder on the dynamics.

The amplitude $A$ of the width oscillations, 
$
A = \left[ {\max_zw(z) - \min_zw(z)} \right]/2,
$
can be computed  by employing the Wigner formalism (details are given in the SM \cite{Supplemental}).
In Fig. \ref{fig3}(d) we plot $A$ against the coupling parameter \( C \) and the initial Gaussian width \( w_0 \).
We also identify the set of parameters for which $A= 0$, corresponding to coherent-state-like evolution.  
These parameters are described by the relation 
\begin{equation}
    w_0 = \left[ C^2 - \left(n_0 - \frac{N+1}{2}  \right)^2 \right]^{1/4},  
    \label{coherent}
\end{equation}
shown as the solid curve.
This condition singles out the initial widths that exactly balance the lattice inhomogeneity, allowing the beam to propagate without undergoing periodic broadening or compression. 
Parameter values below (above) the solid curve lead to a phase (amplitude) squeezing evolution.
These results extend directly to the corresponding two-dimensional lattice, since the evolution along the \( x \)- and \( y \)-directions is decoupled; the overall dynamics can be described as a product of two independent one-dimensional evolutions, each exhibiting  coherent-state-like or squeezing behavior, depending on the respective initial width along that direction. Examples illustrating this extension to 2D are provided in the SM \cite{Supplemental}. Let us note here that up to this point, our analysis has been restricted to linear dynamics.
Next, we explore how  nonlinearity influences the wavepacket evolution.

%###########################################################################

\begin{figure*}
\begin{center}
\includegraphics[width=2\columnwidth]{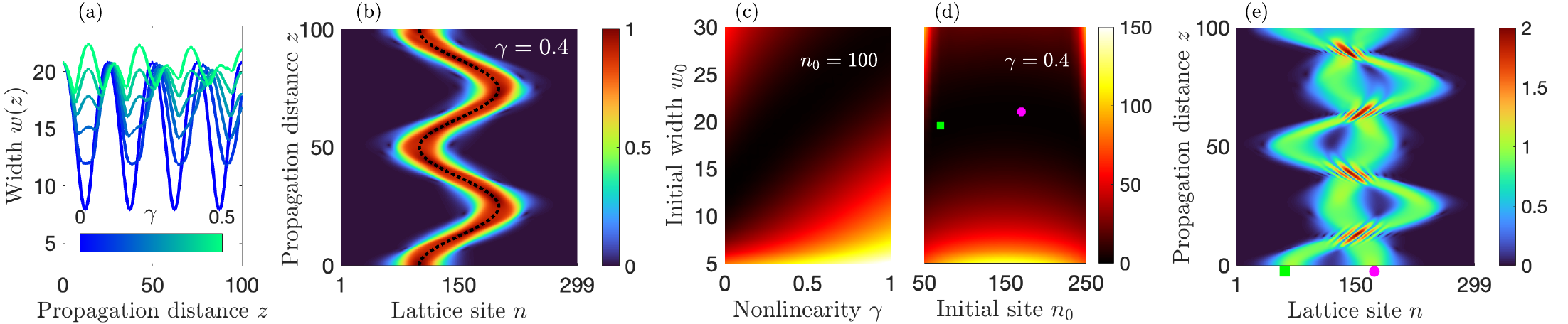}
\caption
{{Impact of Kerr nonlinearity on wave packet evolution.}  
(a) Evolution of the wave packet width for different values of the nonlinearity strength $\gamma$. The initial condition is the same as in Fig. \ref{fig3}(b3).
(b) Evolution of the initial wave packet used in Fig. \ref{fig3}(b3) for $\gamma=0.4$.
(c) Width variance as a function of $(\gamma,w_0)$ for a fixed wave packet center $n_0=100$. Dark regions (minimal variance) correspond to coherent-state-like propagation regimes.
(d) Same as in (c), but showing the width variance as a function of $(n_0,w_0)$ for fixed nonlinearity strength $\gamma=0.4$.
(e) Evolution of two wave packets with zero momentum and parameters $n_0=70$ and $w_0=19.5$ for the first wave packet and $n_0=170$ and $w_0=21$ for the second [these parameters are indicated by the square and circle markers in (d)]. The nonlinearity strength is $\gamma=0.4$.
}
\label{fig4}
\end{center}
\end{figure*}

\PRLSec{Effect of Kerr nonlinearity}
\label{Section4}We now turn to the question of how the Kerr nonlinearity affects the beam dynamics~\cite{Christodoulides1988}.
To this end, we solve the equations $id\psi_n / dz = H_{nm}\psi_m+\gamma |{\psi}_n|^2 \psi_n$ with $n=1,...,N$.
In Fig.~\ref{fig4}(a), we plot the evolution of the beam width for an initially broad Gaussian wave packet [the same as in Fig.~\ref{fig3}(b3)], for five representative values of the nonlinearity parameter \( \gamma \). It is evident that the nonlinearity affects the width oscillations: as \( \gamma \) increases, the oscillation amplitude initially decreases, reaching a minimum at a critical value \( \tilde{\gamma} \). Beyond this point, further increase in \( \gamma \) leads to a renewed growth in the oscillation amplitude. Figure~\ref{fig4}(b) shows the wave packet evolution for a value of \( \gamma \) that results in minimal width fluctuations (\( \gamma = 0.4 \)), illustrating an almost coherent-state-like propagation even in the presence of nonlinearity \cite{Note1}.

The latter findings suggest that a Gaussian wave packet, which propagates in a squeezed way within the linear regime, undergoes coherent-state-like evolution once nonlinearity is introduced, and vice versa. To further explore this transition, we examine the interplay between the initial width $w_0$, the launch point $n_0$, and the nonlinear parameter $\gamma$. Figures \ref{fig4}(c) and \ref{fig4}(d) show the width variance across these parameters, identifying regions where the variance is negligible and the evolution remains effectively coherent-state-like.
Moreover, in Fig. \ref{fig4}(e) we present the evolution of two Gaussian wave packets whose initial parameters are marked with the square and the circle in Fig. \ref{fig4}(d) (these specific parameters correspond to the regime of minimal width variance). We observe that the tautochrone effect persists in the nonlinear regime, as confirmed by the synchronized arrival of the two beams at the lattice center in Fig. \ref{fig4}(e).
Finally, we note that in  SM \cite{Supplemental} we have included additional results regarding self-focusing and perfect-focusing in this type of lattice.

\PRLSec{Conclusions and Discussion}
\label{Conclusions}
In this work we investigate a class of nonuniform photonic lattices that provide an optical analog of the tautochrone problem of classical mechanics and of squeezing dynamics of quantum optics.
More specifically, different initial conditions, lead to coherent-state-like or squeezed  propagation. 
Alternatively, instead  of varying the initial conditions, the Kerr nonlinearity plays a similar role.

These effects are directly realizable based on current experimental setups using femtosecond-laser-written waveguide arrays, where arbitrary coupling profiles can be inscribed with high precision (see for instance Ref.~\cite{Biesenthal2022}). Moreover, the tautochrone focusing and squeezing dynamics could also be relevant in other physical settings, such as ultracold atoms in optical lattices with engineered tunneling rates \cite{Bloch2011, Ketterle2013, Bloch2013} and classical optical thermodynamical setups \cite{Pyrialakos2022, Shapiro2021, Chotorlishvili2025, Wu2019, Wise2022}.  In summary, our work highlights the significance of geometrically engineering the coupling coefficients in discrete optical systems and may provide a new avenue for beam engineering and light control in complex media.

\PRLSec{Acknowledgments}
I.K., G.A.S and K.G.M.  acknowledge support from the European Research Council (ERC-Consolidator) under Grant Agreement No. 101045135 (Beyond\_Anderson) and from the Stavros
Niarchos Foundation (SNF) and the Hellenic Foundation for Research and Innovation (HFRI) through the 5th Call of the Science and Society Action, titled Always Strive for Excellence–Theodoros Papazoglou (Project No. 11496, “PSEUDOTOPPOS”).
G.S. also acknowledges support from the National Natural Science Foundation of China (NSFC) through Grant No. 12474262. 
M.H. and A.S. acknowledge support from the Deutsche Forschungsgemeinschaft (GRK 2676/1-2023 "Imaging of Quantum Systems," project no. 437567992; and SFB 1477 "Light-Matter Interactions at Interfaces," project no. 441234705).

%###########################################################################

% Clear the page and switch to one column
\clearpage
\onecolumngrid

% Reset counters for the Supplement
\setcounter{equation}{0}
\setcounter{figure}{0}
\setcounter{table}{0}
\setcounter{section}{0}
\setcounter{page}{1}

% Prefix Equations, Figures, and Tables with "S"
\renewcommand{\theequation}{S\arabic{equation}}
\renewcommand{\thefigure}{S\arabic{figure}}
\renewcommand{\thesection}{S\arabic{section}}
\renewcommand{\thepage}{S\arabic{page}}

\begin{center}
    \textbf{\large Supplemental Material: Optical Tautochrone and Squeezing Dynamics in Nonuniform Lattices} \\[.5cm]
    Ioannis Kiorpelidis,$^{1,2}$ Matthias Heinrich,$^{3}$ Alexander Szameit,$^{3}$ \\Georgios A. Siviloglou,$^{1,2,4,5}$ and Konstantinos G. Makris$^{1,2}$ \\[.2cm]
    \small\textit{$^1$ITCP-Department of Physics, University of Crete, 71003 Heraklion, Greece\\
    $^2$Institute of Electronic Structure and Laser (IESL), FORTH, 71110 Heraklion, Greece\\
    $^3$Institute for Physics, University of Rostock, Rostock, Germany\\
    $^4$Department of Physics, Southern University of Science and Technology, Shenzhen 518055, China\\
    $^5$International Shenzhen Quantum Academy, Shenzhen 518055, China}
\end{center}

\vspace{0.5cm}

% --- Supplement Content Starts Here ---

% %#################################################################################################

\section{I. Analytic derivation of the coupling scheme}
In this section, we provide the analytic derivation of the inhomogeneous coupling distribution required to realize the optical tautochrone effect. The tight-binding Hamiltonian that governs the dynamics is  
\begin{equation}
{H} = -\sum_{i=1}^{N-1} J_i \left( |i\rangle\langle i+1| + |i+1\rangle\langle i| \right)
\end{equation}
where $|i\rangle$ is the basis state localized at the $i$-th waveguide.

To determine the wave packet trajectory, we consider the continuous  limit: we define a continuous wavefunction $\psi(x)$ such that $\psi(n) = \psi_n$ at integer lattice positions, and let the coupling be described by a smooth function $J(x)$. By utilizing the finite translation operator $e^{\pm i {p}}$, where ${p} = -i\partial_x$ is the momentum operator, we express the neighboring site amplitudes as $\psi(x \pm 1) = e^{\pm i {p}} \psi(x)$ \cite{Brink2021}. Under this approximation, the Hamiltonian operator becomes
\begin{equation}
{H} = -J(x) \left( e^{-i {p}} + e^{i {p}} \right) = -2J(x) \cos p .
\end{equation}

The evolution of the mean position $\langle x \rangle$ and mean momentum $\langle p \rangle$ of a Gaussian wave packet can be determined via the Ehrenfest equations \cite{Longhi2009}
\begin{align}
\frac{d\langle x \rangle}{dz} &= \left[\langle x \rangle, H\right] = 2J(x) \sin p, \label{eq:dxdt_supp} \\
\frac{d\langle p \rangle}{dz} &= \left[\langle p \rangle, H\right] = 2 \frac{dJ(x)}{dx} \cos p. \label{eq:dpdt_supp}
\end{align}
Combining Eqs.~\eqref{eq:dxdt_supp} and \eqref{eq:dpdt_supp}, we obtain the following second-order differential equation for the mean position
\begin{equation}
\frac{d^2  \langle x \rangle }{dz^2} = 4 J(x) \frac{dJ(x)}{dx}. \label{eq:xsecond}
\end{equation}
To achieve the tautochrone effect, the wave packet must perform harmonic oscillations 
because a fundamental property of harmonic oscillators is isochronicity: the time (propagation distance $z$ in the considered lattice) required to reach the equilibrium position is independent of the initial displacement.
This requires that the restoring force in Eq. \eqref{eq:xsecond} is linear with $x$. We therefore impose that the right hand side of Eq.~\eqref{eq:xsecond} is equal to $-4\omega^2(\langle x \rangle - N/2)$ where 
 $\omega$ is the spatial oscillation frequency. This requirement leads to the coupling scheme described in Eq.~(4) of the main text.
Moreover, the analytic solution for the mean position $\langle x \rangle$ can be expressed as
\begin{equation}
\langle x \rangle  = \tilde{x} + ( x_0  - \tilde{x}) \cos(2 \omega z) + \frac{f_{x_0}}{2 \omega} \sin(2 \omega z),
\label{eq:x_analytic}
\end{equation}
where \( \tilde{x} = (N+1)/2 \),  \( f_{x_0} = 2 J(x_0) \sin  p_0  \) and $x_0$ ($p_0$) is the initial position (momentum) of the wave packet.
Namely, the mean position undergoes an oscillatory motion around the lattice center  with frequency $2\omega$.
It is noted here that throughout the main text and the following sections of the SM, we omit the expectation value brackets for reasons of notation convenience. Therefore, $x$  denotes the expectation value of the position operator, or in other words the mean value of the wavepacket's position.

\section{II. Limits of validity of Ehrenfest effective equations}
Here, we examine the limits of validity of the effective continuous Ehrenfest equations of the previous Section I. In particular, in Fig. S1(a)-(c), we compare the numerically calculated trajectory of a beam's center-of-mass (solid black) with the prediction of the Ehrenfest theorem (dashed cyan) in lattices with sizes $N=51$ (panel a), $N=75$ (panel b), and $N=199$ (panel c). For these configurations, we set $C=2N/3$ and launch the beam from $n_{0}=(N+1)/4$.
In Fig. S1(d), we display the difference $|x_{numeric}-x_{Ehrenfest}|$ for these three lattice sizes as a function of propagation distance $z$. We observe that the error oscillates but its magnitude decreases  as the lattice size $N$ increases, confirming that the continuous  approximation becomes more accurate in larger lattices.
Finally, Fig. S1(e) displays the maximum deviation over a propagation distance $z \in [0, 200]$: for sufficiently large $N$ the maximum error remains below 0.5, showing that the difference between the discrete evolution and the continuous  approximation is at a sub-lattice scale.

\begin{figure}[H]
\begin{center}
\includegraphics[width=1\columnwidth]{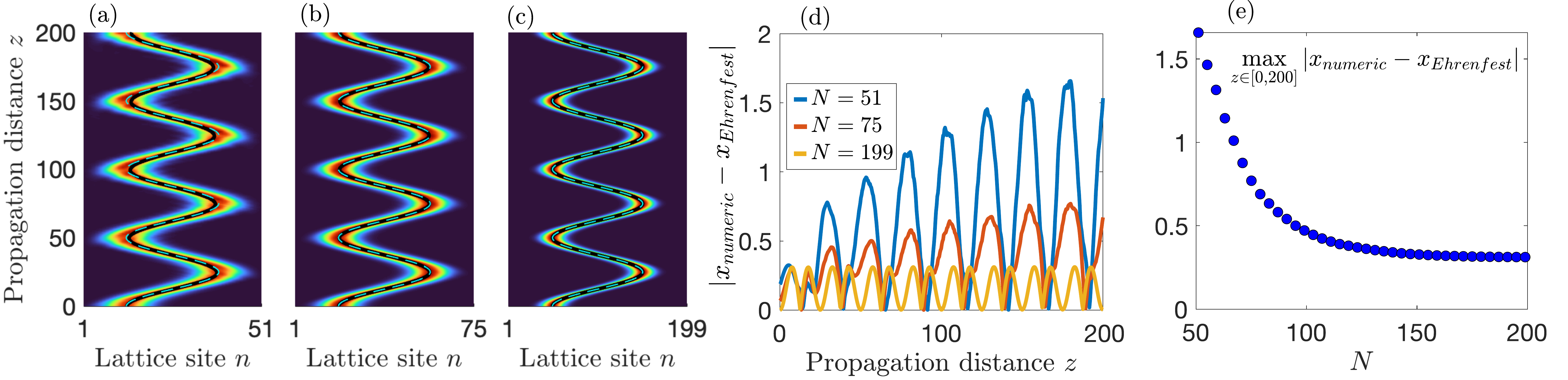}
\caption{(a)-(c) Intensity evolution and center-of-mass trajectories for $N=51, 75, 199$. Solid black lines show numerical results; dashed cyan lines represent Ehrenfest predictions. Parameters: $C=2N/3$, $n_{0}=(N+1)/4$, and $\omega=2\pi/100$. (d) Absolute error $|x_{numeric}-x_{Ehrenfest}|$ versus propagation distance $z$. (e) Maximum deviation for $z\in[0,200]$ as a function of $N$.}
\label{figsupple1n}
\end{center}
\end{figure}

\section{III. Spectrum}
\begin{figure}[H]
\begin{center}
\includegraphics[width=0.5\columnwidth]{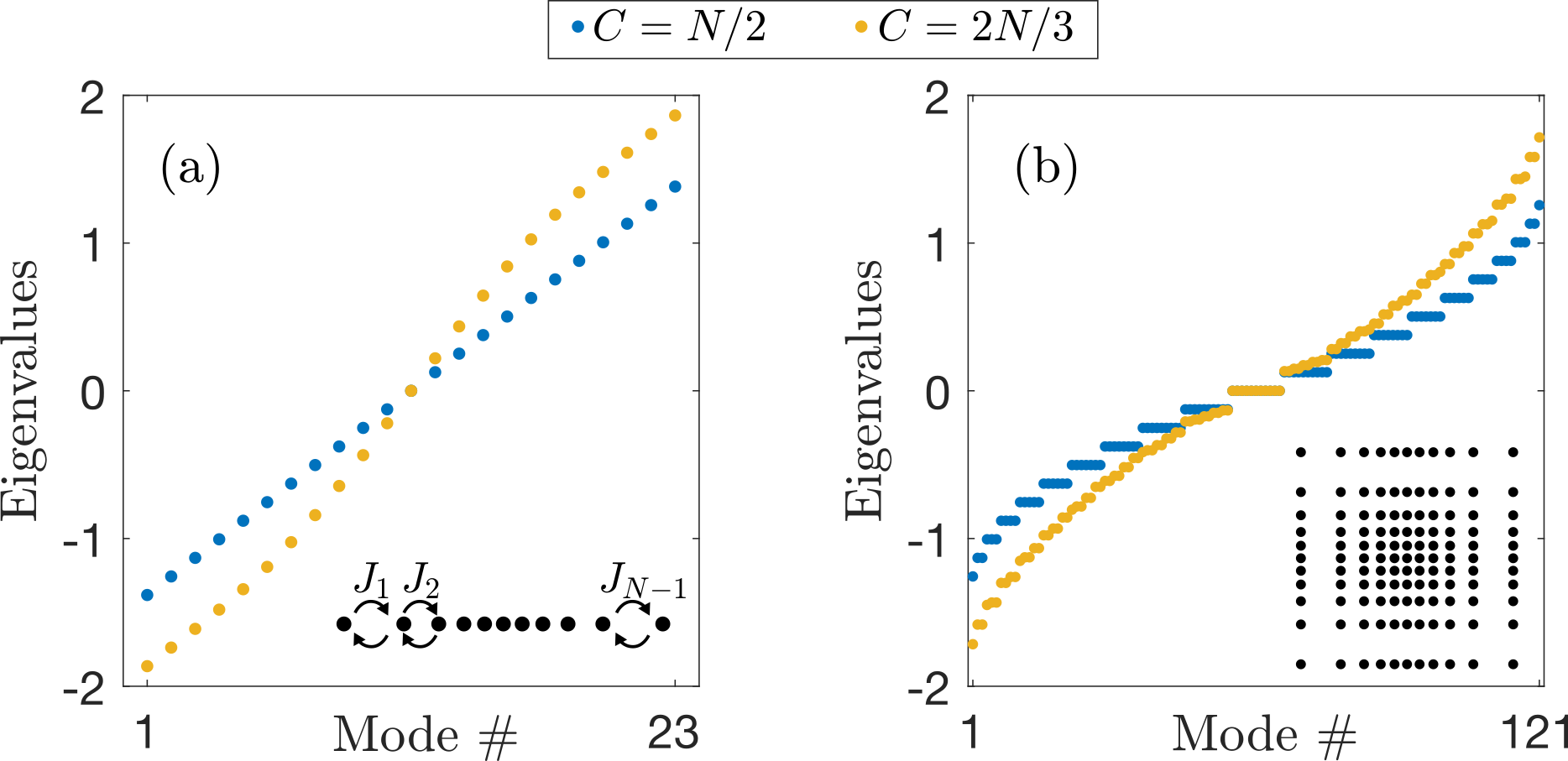}
\caption
{
(a) Eigenvalues for a 1D lattice ($N=23$) with couplings $J_n$ given from Eq. (4) of the main text. Two $C$ values are considered ($C=N/2$ and $C=2N/3$) and $\omega = 2\pi/100$. (b) Eigenvalue spectrum for the corresponding 2D lattice ($N=11 \times 11$).
In both panels (a) and (b), the insets show cross-section schematics of the corresponding waveguide arrays.
}
\label{figsupple1}
\end{center}
\end{figure}

We show here the eigenvalues of the Hamiltonian \( \mathbf{H} \) governing the lattice dynamics.
Figure \ref{figsupple1}(a) displays the 1D spectrum for two values of $C$: the eigenvalues are perfectly equidistant for $C=N/2$ and become non-equidistant for $C>N/2$. Yet, even in the latter case, the spacings at the spectral edges remain nearly uniform, allowing wave packets to maintain almost perfect oscillations even with tailored coupling inhomogeneity.

Figure \ref{figsupple1}(b) shows the corresponding eigenvalue spectrum of the 2D Hamiltonian for \( C = N/2 \) and \( C > N/2 \). Notice first that degeneracies appear in the spectrum, due to the underlying symmetries. Moreover, when \( C = N/2 \), the eigenvalues are commensurate—their differences are integer multiples of a fundamental frequency—resulting in perfect revivals of the wave packet. For \( C > N/2 \) this commensurability is lost, yet the wave packet continues to exhibit {almost perfect} oscillations, similar to the one-dimensional case.

\section{IV. Influence of the parameter $C$ to the tautochrone effect}
The emergence of the tautochrone effect is fundamentally rooted in harmonic oscillating motion. As we demonstrated via the Ehrenfest theorem in Section I of the SM,  a beam’s center of mass undergoes harmonic oscillation within the lattice for any value of $C \ge N/2$. While the continuous  approximate approach provides the theoretical foundation of our analysis, we must account for the fact that our physical system is a discrete lattice and the structure of the eigenvalue spectrum plays a vital role. As shown in Section III of the SM, the eigenvalues are equidistant (non-equidistant) for $C = N/2$ ($C > N/2$). Consequently, when $C = N/2$, the beam trajectories display \textit{perfect revivals}, whereas for $C > N/2$, the trajectories exhibit \textit{almost perfect revivals}. The larger the value of $C$, the greater the deviation from a perfect revival. This suggests that an \textit{exact tautochrone} occurs for $C = N/2$, while for $C > N/2$ the system supports an \textit{almost exact tautochrone}. This behavior is illustrated in Fig. S3, which displays the evolution of three beams for $C = N$ (panel a), $C = 5N$ (panel b), and $C = 15N$ (panel c). In each case, the left sub-panels correspond to noninterfering beams, while the right sub-panels correspond to interfering beams.

\begin{figure}[H]
\begin{center}
\includegraphics[width=1\columnwidth]{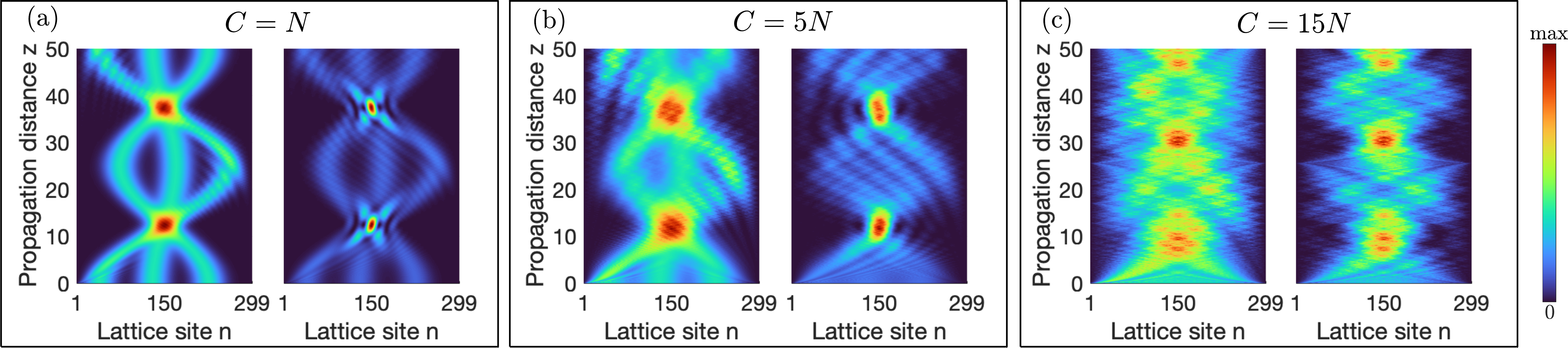}
\caption
{Intensity evolution of three Gaussian wave packets in a 1D lattice ($N=299$ sites). All parameters are the same as in Figs. 2(a1) and 2(a2) of the main text, except for the inhomogeneity parameter $C$. (a) $C = N$, (b) $C = 5N$, and (c) $C = 15N$. In each panel, the left sub-panels depict the evolution of noninterfering beams as in Fig. 2(a1) of the main text, while the right sub-panels correspond to interfering beams as in Fig. 2(a2)  of the main text.}
\label{figsupple3n}
\end{center}
\end{figure}

\section{V. Focusing of random initial states}
\begin{figure}[H]
\begin{center}
\includegraphics[width=0.55\columnwidth]{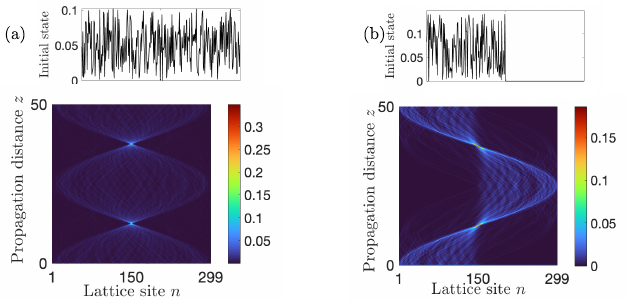}
\caption{(a) Evolution of the site-resolved intensity (lower panel) of a random initial state (upper panel) that is extended across the entire lattice ($N=299$ sites). (b) Same as (a), but the random initial state (upper panel) is extended only over the left half of the lattice.}
\label{supple3}
\end{center}
\end{figure}

We show here the evolution of a random initial state that is extended across the whole lattice [Fig.~\ref{supple3}(a)] and over half of the lattice [Fig.~\ref{supple3}(b)]. In both cases, the state is localized in momentum space. In panel Fig.~\ref{supple3}(a), the evolution is similar to that of a breathing mode and focuses at the lattice center at a propagation distance $z = \pi / (4\omega)$. In Fig.~\ref{supple3}(b), the state focuses again at the same propagation distance $z = \pi / (4\omega)$, yet the center of mass traces out an oscillatory motion.

\section{VI. Impact of Coupling Disorder}

\begin{figure}[H]
\begin{center}
\includegraphics[width=1\columnwidth]{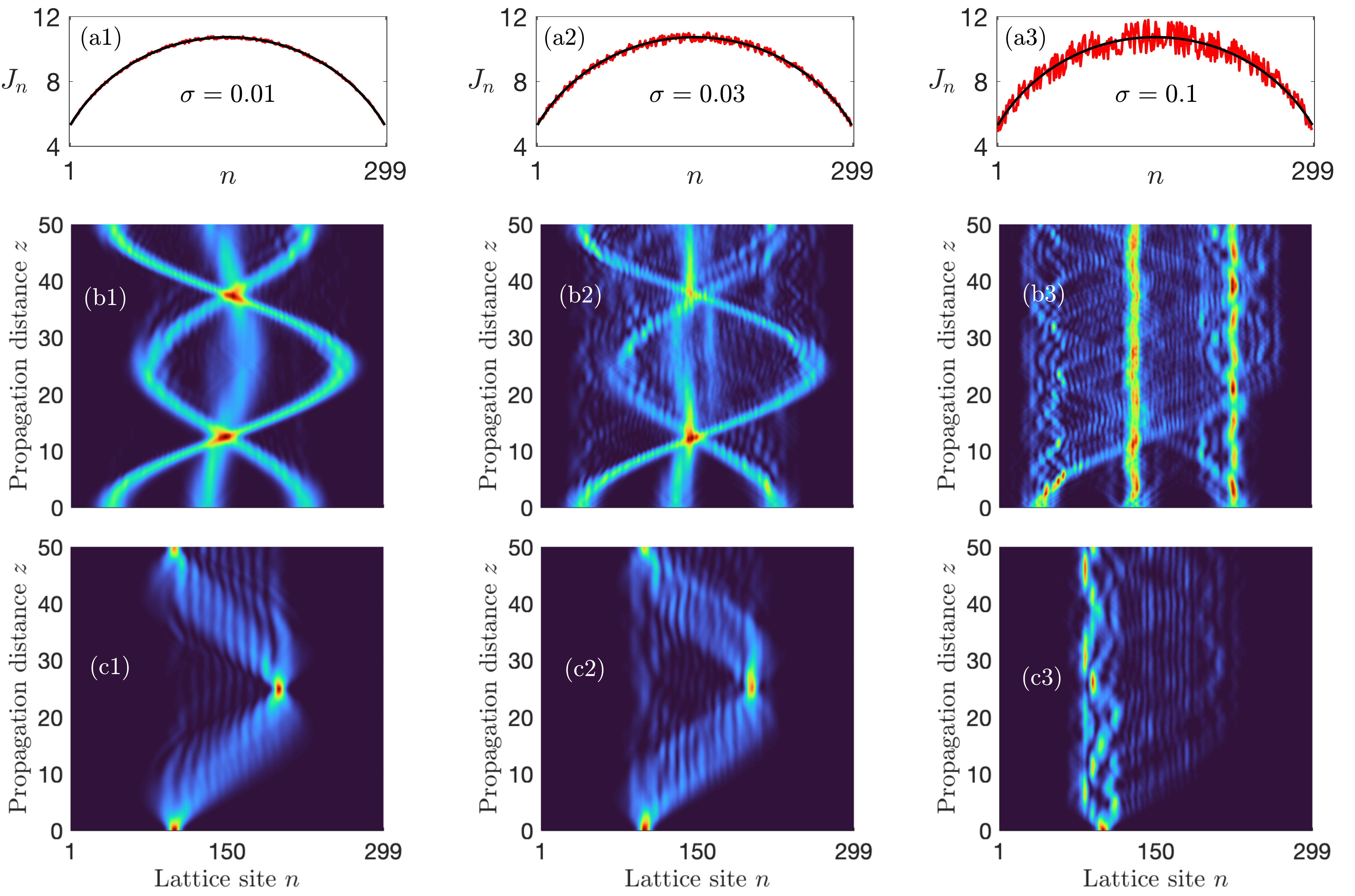}
\caption
{(a1)–(a3) Coupling profiles $J_n$ for disorder strengths $\sigma = 0.01, 0.03,$ and $0.1$. (b1)–(b3) Intensity evolution of three noninterfering Gaussian wave packets. All parameters are the same as in Fig. 2(a1) of the main text, except for the introduction of disorder. (c1)–(c3) Corresponding evolution of a narrow Gaussian input. All parameters are the same as in Fig. 3(b1) of the main text.}
\label{figsupple4n}
\end{center}
\end{figure}

To investigate the robustness of the reported phenomena, we perform numerical simulations by considering random disorder in the coupling coefficients. Specifically, we modulate the hopping amplitudes according to:
$$J_{n}^{\text{disordered}} = J_{n}(1 + \sigma \epsilon)$$
where $\sigma$ represents the disorder strength and $\epsilon$ is a random variable uniformly distributed in the interval $[-1, 1]$. We consider three representative cases for the disorder strength: $\sigma = 0.01$ (left panels of Fig.~\ref{figsupple4n}), $\sigma = 0.03$ (middle panels of Fig.~\ref{figsupple4n}), and $\sigma = 0.1$ (right panels of Fig.~\ref{figsupple4n}).

As illustrated in Fig.~\ref{figsupple4n}, for low to moderate disorder levels ($\sigma = 0.01$ and $\sigma = 0.03$), the tautochrone (panels b1-b2) and squeezing (panels c1-c2) effects persist despite the broken symmetry of the lattice. For high disorder strength ($\sigma = 0.1$), the tautochrone and squeezing effects are naturally suppressed because the strong random fluctuations lead to Anderson localization. In this regime, the waves become spatially trapped near their initial positions, preventing the beams from reaching the lattice center.

\section{VII. Evolution of the Wave packet Width}
We give here the derivation of the  evolution of the wave packet width.
The initial state is described by a Gaussian distribution in phase space of the form \cite{Mossmann2004}
\begin{equation}
    W(x, p, 0) = \frac{1}{2\pi} \exp\left(-\frac{(x - n_0)^2}{2 w_0^2} - \frac{p^2 w_0^2}{2}\right).
\end{equation}
To compute the  evolution of the wave packet width, we evaluate the variance\( \Delta x^2(z) = \overline{x^2}(z)  -  \overline{x}^2(z) \), where the averages are taken over the initial Gaussian distribution in phase space. Introducing \( u = n_0 - \tilde{x} \), the squared width takes the form:
\begin{equation}
\begin{split}
\Delta x^2(z) =\, & \Delta x_0^2 \cos^2(2\omega z) + \sin^2(2\omega z)   \bigg[ (C^2 - u^2)\cos^2 p \,\Delta p_0^2 
+ \frac{u^2 \sin^2 p}{C^2 - u^2} \Delta x_0^2 \bigg] 
 - 2 \cos(2\omega z) \sin(2\omega z) 
\cdot \frac{u \sin p}{\sqrt{C^2 - u^2}} \Delta x_0^2.
\end{split}
\label{eq:width_general_corrected}
\end{equation}
% This expression shows that the width \(\Delta x(t)\) undergoes oscillations with frequency \(2\omega\), whose amplitude depends on both the initial offset \(u\) and momentum \(p\).

For zero initial momentum, \( p = 0 \), the expression simplifies. In this case, the cross term vanishes and the squared width becomes
\begin{equation}
\Delta x^2(z) = R + G \cos(4\omega z),
\end{equation}
where
\begin{align}
R &= \frac{1}{2} \left( \Delta x_0^2 + {(C^2 -  u^2)} \Delta p_0^2 \right), \\
G &= \frac{1}{2} \left( \Delta x_0^2 - {(C^2 -  u^2)} \Delta p_0^2 \right).
\end{align}
Consequently, the width exhibits simple harmonic oscillations at frequency \(4\omega\) with amplitude
\begin{equation}
A = \frac{1}{2} \left| \sqrt{R + G} - \sqrt{R - G} \right|. \label{eq:amplitude}
\end{equation}

From Eq.~\eqref{eq:amplitude}, the oscillation amplitude vanishes (i.e., the width becomes constant in time) if and only if \( G = 0 \). This leads to the condition:
\begin{equation}
\Delta x_0^2 = (C^2 -  u^2) \Delta p_0^2.
\end{equation}
Substituting \( \Delta x_0 = w_0 \) and \( \Delta p_0 = 1 / w_0 \), we find the relation given in Eq.~(7) of the main text; the condition leading to coherent-state-like propagation.

\section{VIII. Squeezing evolution in the two-dimensional lattice}
\begin{figure}[H]
\begin{center}
\includegraphics[width=0.9\columnwidth]{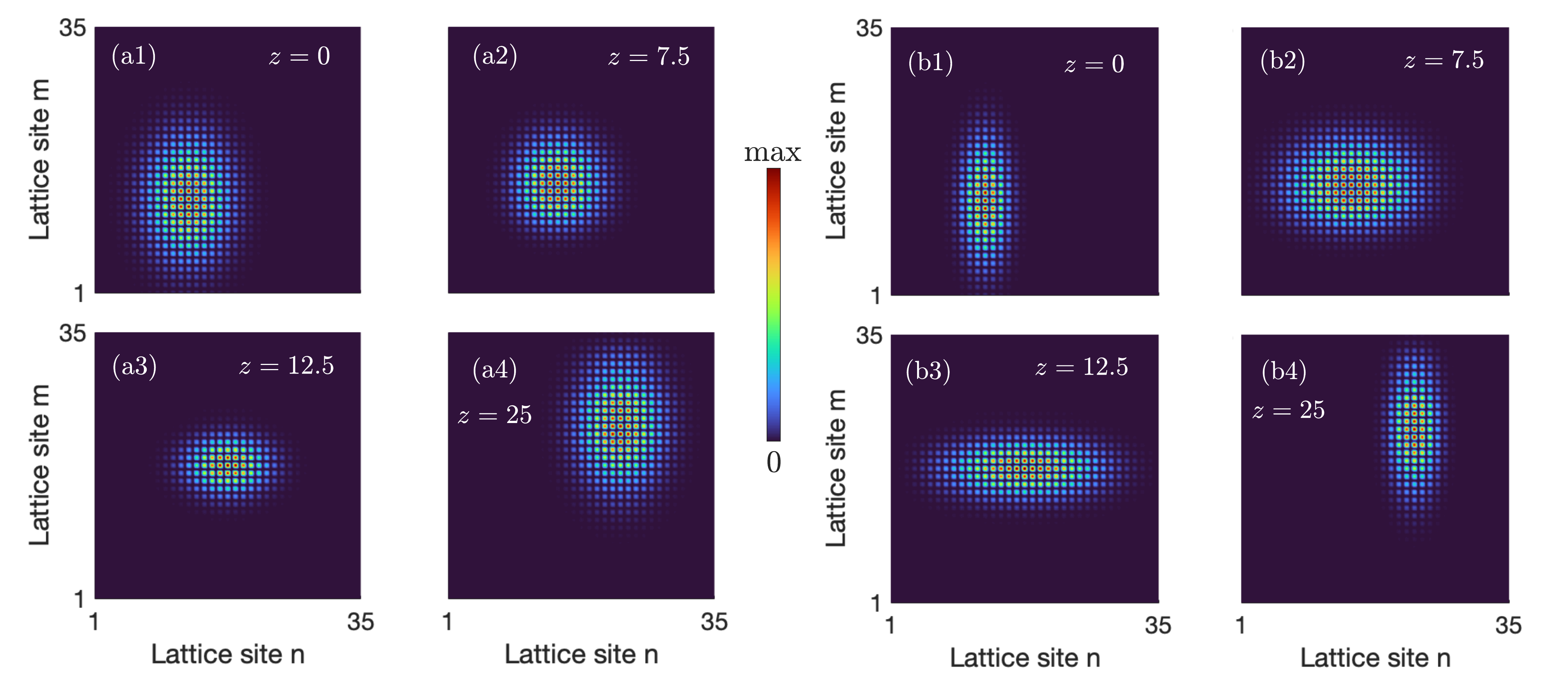}
\caption
{
(a1)–(a4) Evolution of a Gaussian wave packet centered at \( (n_0 = 13, m_0 = 13) \), with initial widths \( w_{0,x} \approx 4.4 \) (coherent width) along the \( x \)-direction and \( w_{0,y} \approx 6.4 \) along the \( y \)-direction. The wave packet has zero initial momentum. The lattice features equal coupling profiles and frequencies along both axes, with parameters \( C = \tfrac{4N}{7} \) and \( \omega = \tfrac{2\pi}{100} \).
(b1)–(b4) Same as in (a1)–(a4), but the wave packet has width \( w_0 \approx 2.4 \) along the  \( x \) and \( w_{0,y} \approx 6.4 \)  along the  \( y \) directions.
}
\label{supple2}
\end{center}
\end{figure}

We present here two illustrative examples of squeezed evolution in the two-dimensional nonuniform photonic lattice.
It is firstly stressed that for illustration reasons, in the figures of the main text and of the SM we omitted the regions between the waveguides, while in Fig.~\ref{supple2} we present the propagation including these regions.
In particular, in Figs.~\ref{supple2}(a1)–(a4) we display four snapshots of the propagation of an initial Gaussian wave packet whose width exceeds the coherent width in the \( y \)-direction while being equal to the coherent width in the \( x \)-direction. The beam undergoes periodic squeezing in the \( y \)-direction, while its transverse profile remains unchanged along \( x \).
Figures~\ref{supple2}(b1)–(b4) show the evolution for a different initial condition, where the width of the input beam is again larger than the coherent width in the \( y \)-direction, but now smaller than the coherent width in the \( x \)-direction. In this case, the beam experiences squeezing in both directions, with its shape periodically expanding and contracting during propagation.

\section{IX. Relation to Self-Focusing and Perfect-Focusing via Phase Conjugation}

\begin{figure}[H]
\begin{center}
\includegraphics[width=1\columnwidth]{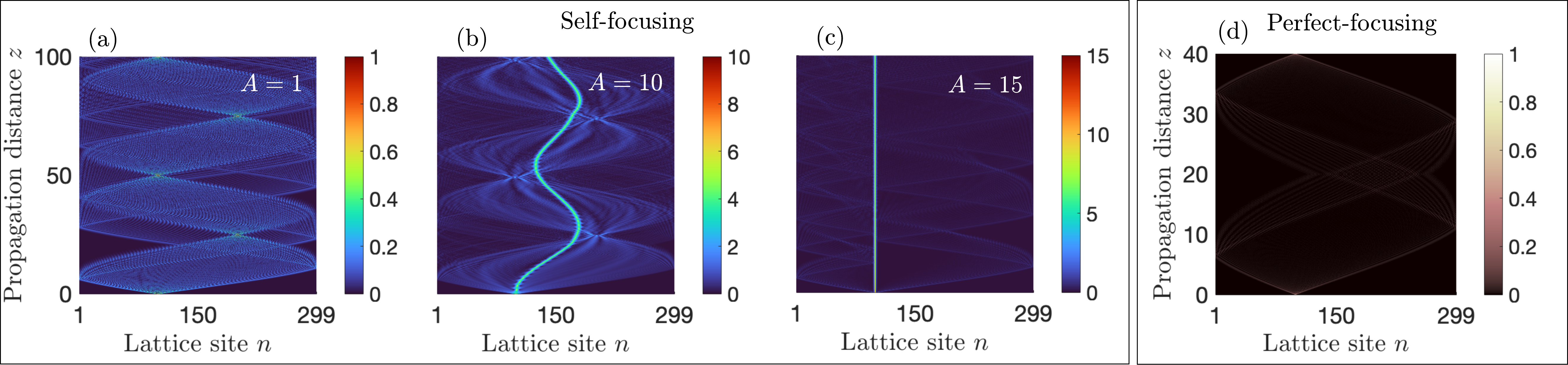}
\caption
{(a)-(c) Intensity evolution at $n_0=100$ with nonlinearity $\gamma=0.4$. Increasing the amplitude $A$ leads to spatial trapping (self-focusing). (d) Perfect focusing in a linear lattice via phase conjugation at $z=20$, showing the field returning to its original single-site state at $z=40$.}
\label{supple7}
\end{center}
\end{figure}

In this section, we examine self-focusing and perfect-focusing effects in our nonunifrom lattice.
Regarding self-focusing, we first note that we introduce Kerr nonlinearity  to the system.  In Figs. \ref{supple7} (a)-(c), we present the evolution of a single-site excitation launched at $n_0=100$.
Figure \ref{supple7} (a) shows the evolution with an initial amplitude of $A=1$. Figure \ref{supple7}(b) demonstrates the evolution as the initial amplitude is increased to $A=10$. Figure \ref{supple7}(c) reveals that at a high initial amplitude ($A=15$), the excitation becomes self-trapped, clearly demonstrating self-focusing within this nonuniform system.

Regarding perfect-focusing, we note that the considered linear lattice is a conservative (Hermitian) system. In such systems, perfect focusing can  be achieved through time-reversal symmetry or back-propagation by employing phase conjugation. In Fig. \ref{supple7} (d), we demonstrate this mechanism where a localized input at $n_{0}=100$ is propagated forward for a distance $z_{max}=20$. At this point, we take the complex conjugate of the optical field ($\psi \rightarrow \psi^{*}$), which effectively reverses the phase. The conjugated field is then propagated for an additional distance $z_{max}=20$, and due to the Hermitian nature of the Hamiltonian, the field retraces its path and focuses perfectly back to its original single-site state at $z=40$.

\end{document}